\documentclass{article}
\usepackage{slashed}

\def\beq{\begin{eqnarray}}
\def\eeq{\end{eqnarray}}
\def\bea{\begin{eqnarray*}}
\def\eea{\end{eqnarray*}}

\def\centeron#1#2{{\setbox0=\hbox{#1}\setbox1=\hbox{#2}\ifdim
\wd1>\wd0\kern.5\wd1\kern-.5\wd0\fi
\copy0\kern-.5\wd0\kern-.5\wd1\copy1\ifdim\wd0>\wd1
\kern.5\wd0\kern-.5\wd1\fi}}
\def\ltap{\;\centeron{\raise.35ex\hbox{$<$}}{\lower.65ex\hbox{$\sim$}}\;}
\def\gtap{\;\centeron{\raise.35ex\hbox{$>$}}{\lower.65ex\hbox{$\sim$}}\;}

\def\singleandthirdspaced{\baselineskip=\normalbaselineskip\multiply
    \baselineskip by 130\divide\baselineskip by 100}

\newcommand{\newc}{\newcommand}
\newc{\qbar}{{\overline q}}
\newc{\Kahler}{K\"ahler }
\newc{\deltaGS}{\delta_{\rm GS}}

\begin{document}
\begin{titlepage}
\begin{flushright}
{\large hep-th/yymmnnn \\
SCIPP 11/03\\
}
\end{flushright}

\vskip 1.2cm

\begin{center}

{\LARGE\bf Studies in Small Field Inflation}

\vskip 1.4cm

{\large  Michael Dine and Lawrence Pack}
\\
\vskip 0.4cm
{\it Santa Cruz Institute for Particle Physics and
\\ Department of Physics,
     Santa Cruz CA 95064  } \\
\vskip 4pt

\vskip 1.5cm

\begin{abstract}
We explore some issues in slow roll inflation in situations where field excursions are
small compared to $M_p$.  We argue that for small field inflation, minimizing fine tuning requires low energy
supersymmetry and a tightly constrained structure.  Hybrid inflation
is {\it almost} an inevitable outcome.  The resulting theory can be described in terms of a supersymmetric low energy
  effective action and inflation completely characterized in terms of a small number of parameters.  Demanding slow roll inflation significantly
  constrains these parameters. In this context, the
  generic level of fine tuning can be described as a function of the number of light fields, there is an upper bound on the scale of inflation, and an (almost) universal prediction
  for the spectral index.  Models of this type need not suffer from a cosmological moduli problem.
\end{abstract}

\end{center}

\vskip 1.0 cm

\end{titlepage}
\setcounter{footnote}{0} \setcounter{page}{2}
\setcounter{section}{0} \setcounter{subsection}{0}
\setcounter{subsubsection}{0}

\singleandthirdspaced

\section{Introduction}
\label{introduction}

There is good evidence that the universe underwent a period of inflation early in its history.
Yet it is probably fair to say that there do not exist completely
reliable, calculable microscopic theories of inflation.
Slow roll inflation provides a simple phenomenology; many, if not
most, microscopic scenarios for inflation, involving branes, extra dimensions and the like, admit such a description.  Indeed, slow
roll inflation makes clear why it is nearly impossible, at present, to formulate
a compelling microscopic theory.  Planck scale effects are {\it necessarily} important, and this requires a full understanding of issues
like dynamics of moduli and supersymmetry breaking, even within a consistent theory of gravity (i.e. a string model).  Even as a phenomenology,
there are a number of models for slow roll inflation.  We follow the review of Lyth\cite{lyth} in dividing these into ``large field", ``medium field" and ``small field" types, where
large, medium or small here refers to field variations much larger than, comparable, or much smaller than the Planck mass.

Almost by definition, large or medium field inflation is difficult to describe in a systematic fashion, without a complete theory of quantum gravity.  Small field inflation,
however, is another matter.  Here one should be able to characterize inflaton in terms of a low energy effective action for some number of light fields, with a
limited set of relevant parameters.\footnote{Expansions of this type have been considered by
various authors; an early discusison appears in \cite{hybrid3}.}$^,$\footnote{Here we are not using ``relevant" in the conventional renormalization group sense;
but instead referring to their relevance to inflation; the correspondence to the usual terminology will be clear shortly.}  This is the goal
of the present paper.  We will see that with some very mild assumptions about genericity, we can characterize small field inflation quite simply:
\begin{enumerate}
\item
The effective theory should exhibit an approximate (global) supersymmetry.  Otherwise, the theory is {\it extremely} tuned.
\item
The effective theory should obey a discrete $R$ symmetry.  This accounts for smallness of the superpotential during inflation, the absence
of terms in the effective action which would spoil inflation, and leads to an approximate $R$ symmetry which accounts for supersymmetry
breaking.  While many models of inflation posit a continuous $R$ symmetry, such symmetries almost certainly don't exist in
consistent theories of quantum gravity.  
\item
Supersymmetry is spontaneously broken in the effective theory which describes inflation.  This breaking is not related in any simple
way to the breaking of supersymmetry in the universe at present and the scale is far higher than the TeV scale.
\item
The (approximate) goldstino may or may not lie in a multiplet with the
inflaton.
\item  The effective theory exhibits an approximate, continuous $R$ symmetry, as an accidental (but typical) consequence of the discrete $R$ symmetry.
\item   The continuous $R$ symmetry is not exact; terms allowed by the discrete symmetry break the continuous global symmetry
and spoil inflation, unless the inflationary scale (the square of the Goldstino
decay constant) is sufficiently small.  In other words, there is an upper bound on the scale of inflation.
\item  If the requirement above is satisfied, there are further requirements on the Kahler potential in order to obtain slow roll inflation with
adequate $e$-foldings.  This sets an {\it irreducible} minimum amount of fine tuning necessary to achieve acceptable inflation.  This tuning
grows in severity with the number of Hubble mass fields.
\item  In order that inflation ends, the inflaton must couple to other light degrees of freedom, or must have appreciable self-couplings
in the final ground state.  The coupling to this extra field, or the self couplings, are fixed by $\delta \rho \over \rho$ and the inflationary scale.  In the case of an extra
 field, the resulting structure is necessarily what is called ``hybrid inflation"\cite{lindehybrid, hybrid1,hybrid2,hybrid3,linderiotto}.  In the latter, which we will call
 ``R breaking inflation" (RBI), further fine tuning is required.  In either case, the spectral index is less than one.
    \end{enumerate}

Many of these points have been made before, but perhaps not in the systematic fashion discussed here; in particular, the inevitability of this
structure for small field inflation does not seem to be appreciated.  Similarly, most models of hybrid inflation invoke supersymmetry
and $R$ symmetries\cite{linderiotto,hybrid1, hybrid2, hybrid3, shafi1,shafi2,shafi3,fischler}.  However, the approximate supersymmetry of the effective action
 has not been stressed; more important, the $R$ symmetries have generally been taken to be continuous, and the consequences of discreteness, particularly the upper bound on the
scale of inflation, have not been considered before, to our knowledge.  The restrictions on the Kahler potential have been noted in early work\cite{hybrid1,linderiotto}, but
then seem frequently ignored; their role in determining the irreducible level of fine tuning and the possible number of
$e$-foldings, particularly sharp in light of the observations about $R$ symmetry, does not seem to have been
appreciated.

Given the role of supersymmetry in small field inflation, it is natural to investigate embedding this structure into theories of low energy breaking.  This has the potential
to expose connections between scales of inflation and scales of low energy physics, as we discuss.  Ideas concerning metastable, dynamical supersymmetry breaking
raise the prospect of sharpening these connections further.

In the following sections, we elaborate on each of these points.  In section \ref{inflationandsusy}, we explain the requirement for supersymmetry and the structure of the effective action.
In sections \ref{effectiveaction},\ref{kahlerconstraints}, we relate the inflationary observables to the parameters of the effective action.
In section \ref{lightfields}, we determine how the level of fine tuning depends on the number of fields.  In section \ref{wrconstraints}, we obtain
an upper bound on the scale of inflation from interactions which violate the continuous $R$ symmetry.  Section \ref{reheating} discusses the problem of reheating.
In section \ref{nonhybrid}, we consider possible alternative structures which might result from relaxing our assumptions; non-hybrid models can arise
but require some additional fine tuning and/or degrees of freedom.  In sections \ref{susybreaking},\ref{retrofitting},  we extend the inflationary models to generate low energy supersymmetry
breaking, in the latter section dynamically.  In our concluding section,  we briefly consider
 large field inflation and provide an assessment.

\section{Inflation as Spontaneous Supersymmetry Breaking}
\label{inflationandsusy}

If we impose some fine tuning or genericity constraints, we can characterize small field inflation quite simply.  First we require a field which is light compared to
the Hubble constant during inflation, $H_0$. Scalar fields with mass {\it of order} $H_0$ are already
unnatural, unless the theory, at least at these scales, is supersymmetric\footnote{Ordinary dynamical symmetry
breaking is problematic, since one needs a vast mismatch between the associated energy scale and the mass scale
of the excitations; Goldstone excitations, as in ``natural" inflation, require decay constants much greater than the Planck scale.}.  As we will review, scalar fields with mass much smaller than $H_0$, even with
supersymmetry, are unnatural.  So having one field light compared to $H_0$ represents a fine tuning; it would seem unlikely
that there is more than one such field.  For $H_0 \ll M_p$, we can write a supergravity effective action in an approximately flat space.
More precisely, given our assumption that the inflaton is small compared to $M_p$, the system is described, approximately,
by a globally supersymmetric effective action which exhibits {\it spontaneous} supersymmetry breaking.
This may seem obvious, but it is perhaps worth pointing out that it follows from the assumption of small field excursions.
Consider, for simplicity, a single light field, $S$.  For fields small compared to $M_p$, we can write:
\beq
K \approx K_0 + S^\dagger S + \alpha {(S^\dagger S)^2 \over M_p^2} + \dots
\eeq
Similarly, we can expand $W$ in a power series in $S$.
\beq
W = W_0 + \mu^2 S + {m \over 2} S^2 + {\lambda \over 3} S^3 + \dots.
\eeq
Because of the small field assumption, $W_0$ cannot dominate during inflation,
so
\beq
W_0 < H_0 M_p^2.
\eeq
Then $\mu^2 \sim H_0 M_p$.  The slow roll conditions then imply
\beq
W_0 \ll \mu^2 M_p;~~ m \ll \mu \left ( {\mu \over M_p} \right ); ~~\lambda \ll {\mu^2 \over M_p^2}.
\eeq
There are also constraints on the Kahler potential parameter $\alpha$ which we will discuss shortly.
Formulated in this way, the fermionic component of $S$ is the Goldstino, and its scalar component is the inflaton. The possibility that there is another chiral
field, whose scalar component is the inflaton will be considered further when we discuss the Kahler potential constraints.

The absence (smallness) of terms $W_0$, $S^2,~S^3$ is most readily
accounted for if the theory possesses an $R$ symmetry.  A constant in the superpotential
can only be forbidden by an $R$ symmetry.  Powers of $S$ might be accounted for if $S$ carries a charge,
and $\mu^2$ is a spurion for the corresponding symmetry.   This can be understood as a version of the Nelson-Seiberg theorem\cite{nelsonseiberg}, which requires
an (approximate) continuous $R$ symmetry for (meta-)stable supersymmetry breaking.  Because we do not expect continuous global symmetries,
we will assume that the underlying $R$ symmetry is discrete (and we will, in general, take it to be $Z_N$, $N>2$), while the
continuous symmetry is an accident.  In a moment, we will see that there are further reasons
that an $R$ symmetry seems necessary for successful inflation.

At the end of inflation, supersymmetry must be restored, and the cosmological constant vanish.  One might try to model this without adding
additional degrees of freedom.  Once the $R$ symmetry is understood as a discrete symmetry, one expects
higher order terms in the $S$ superpotential, and supersymmetric vacua at large fields.  We will see that understanding inflation
in terms of flow towards such a minimum is possible, but adds additional complications.  So we will first add additional degrees of freedom coupled to $S$.  For the moment
we will suppose that there is one such field, $\phi$.  Any additional light fields coupled
to $S$ are likely to be quite light at the end of inflation.  $\phi$ gains mass by combining with $S$; if $\phi$ were, say, in a non-trivial representation of
a non-abelian symmetry, only one component could gain mass through this coupling; we will discuss this
issue further later.
We are led, then, to write
\beq
W = S (\kappa \phi^2 - \mu^2) + ~{\rm non-renormalizable~terms}.
\label{standardhybrid}
\eeq
This system has a supersymmetric minimum at $\kappa \phi^2 = \mu^2, ~S=0$.  Classically, however, it has a moduli space with
\beq
\vert \kappa S \vert^2 > \vert \kappa \mu^2 \vert.
\label{pseudomoduli}
\eeq
This pseudomoduli space will be lifted both by radiative corrections in $\kappa$, quantum and Planck-supressed
terms in the Kahler potential, and the non-renormalizable terms in the superpotential.
As we will see, all are necessarily relevant if inflation occurs in the model.   Inflation takes place on this pseudomoduli space;
$\phi$ is effectively pinned at zero during inflation.

The $R$ symmetry now explains the absence of additional dangerous couplings, such as $\phi^2$.  Perhaps most strikingly, though,
it forbids a constant in the superpotential, guaranteeing that at the end of inflation, when supersymmetry is (nearly) restored,
the vacuum energy (nearly) vanishes.  We will explore, in section \ref{nonhybrid}, non-hybrid models in which the superpotential
must be suitably tuned.

Three points should be noted:
\begin{enumerate}
\item
The assumption that the symmetry is discrete
means couplings like $S^{N+1} \over M_p^{N-2}$ are permitted, and, as we will soon see, they significantly constraint inflation.
\item  There are additional conditions, as we will shortly enumerate, on the Kahler potential in order that one obtain inflation with an adequate
number of $e$-foldings.  These constitute at least one fine tuning needed to obtain an inflaton with mass small compared to the Hubble scale.
\item  This structure is not unique; the inflaton need not lie in a supermultiplet with the Goldstino.  If there are several multiplets with
  non-zero $R$ charges, it is possible to tune parameters so that the scalar component of one of these other multiplets
  is light, while the partner of the Goldstino
  is heavy.  As an example, one can contemplate another field, $I$.  If $I$ couples simply to $\phi^2$, this is problematic; inflation
  does not end.  So it is necessary to introduce a field $\phi^\prime$, and take for the superpotential
\beq
W = S(\kappa \phi^2 - \mu^2) +\lambda I \phi \phi^\prime + \dots
\label{modifiedhybrid}
\eeq
Note that at the minimum of the potential (assumed to be at $S = I = \phi^\prime =0$), all fields are massive.
\item  When we consider the constraints arising from points (1) and (2) above, and the values of the cosmological parameters, we will see that the basic
hybrid model of eqn. \ref{standardhybrid} does not produce suitable inflation unless the $N$ of the $Z_N$ is very large.  The model of eqn. \ref{modifiedhybrid}, on the other hand, {\it does} produce
successful inflation for suitable values of the parameters and modest $N$.
\item  We will see that these structures can be embedded in a model of low energy supersymmetry breaking. This is not required, but
would seem elegant and economical.
\end{enumerate}

\section{The Structure of the Effective Action}
\label{effectiveaction}

In this section, we assume supersymmetry at scales above the Hubble constant of inflation, and the presence of a discrete $Z_N$ R-symmetry.
We focus, for now, on the single field model.  Our considerations will generalize
immediately to the multi-field case.
For slow roll, it is crucial that the curvature of the $S$ potential, during inflation, be smaller than the Hubble constant,
\beq
V^{\prime \prime} \ll {\mu^4 \over M_p^2}.
\eeq
This is a strong condition, and as we will describe, requires tuning the parameters of the potential for $S$.  An even stronger
condition arises from the requirement that the field actually flows towards the minimum at the origin.

We have already argued that an $R$ symmetry is a necessary ingredient in successful
small-field inflation.  Because we do not expect
continuous global symmetries in nature, the $R$ symmetry must be discrete; we will take it to be $Z_N$.
So the superpotential has the form of eqn. \ref{standardhybrid}, but with additional terms, which we assume to be Planck suppressed\footnote{Such terms
in inflationary models have been considered in \cite{hybrid4}, where they also play an important role; the scales assumed there, and the detailed picture of
inflation, are quite different, though they resemble some of our discussion in section \ref{nonhybrid}.}:
\beq
W = S(\kappa \phi^2 - \mu^2) + W_R;~~W_R= {\lambda \over 2(N+1) }{S^{N+1} \over M_p^{N-2}} + {\cal O}\left  ({S^{2N + 1} \over M_p^{2N-2}} \right ) .
\label{hybridw}
\eeq
We have called the $S^{N+1}$ term $W_R$ because it breaks the would-be continuous $R$ symmetry.

The Kahler potential is critical to obtaining a mass for $S$ much smaller than the Hubble constant and successful inflation.  Expanding in powers of $S$,
and exploiting the assumption of $R$ symmetry, it takes the form
\beq
K = S^\dagger S + \phi^\dagger \phi- {\alpha \over  4 M_p^2} (S^{\dagger} S)^2 + \dots.
\label{hybridk}
\eeq
We are assuming that, apart from $\mu$, the Planck scale is the only relevant scale; if this is not the case, the fine-tuning problems we discuss
below will be more severe.
With this assumption, the neglected terms will not be important when the fields are small.
We have assumed that other physics is controlled by the Planck scale.

For large values of the fields, the supergravity contributions to the potential,
arising from the quartic terms in the Kahler potential, dominate the potential for $S$.  Most
importantly, they give rise to a quadratic term\cite{hybrid1,linderiotto}:
\beq
V_{SUGRA} = \alpha \mu^4{S^\dagger S \over M_p^2}.
\eeq
For lower values of $S$, the quantum corrections arising from integrating out $\phi$, can be important.
In particular, in the regime where $\vert \kappa^2 S^2 \vert \gg \vert \kappa \mu^2 \vert$ (i.e. on the pseudo moduli
space, eqn. \ref{pseudomoduli}),  
these corrections are easily computed\cite{hybrid2,linderiotto,fischler}:
\beq
\delta K_{quant}(S,S^\dagger) = {\kappa^2 \over 16 \pi^2} S^\dagger S\log(S^\dagger S).
\eeq
This Kahler potential is appropriate to the description of the theory at scales below $\kappa S$, and at times when $S$
is slowly varying.
The corresponding quantum correction to the potential is:
\beq
V_{quant} = {\kappa^2 \over 16 \pi^2} \mu^4 \log(S^\dagger S)
\eeq

The quantum corrections will dominate on the pseudo moduli space if the scale, $S_{quant}$, below which the quantum
contribution to the potential are larger than the supergravity contribution, lies on the moduli space.  $S_{quant}$ is obtained
by comparing the second derivative of $V_{SUGRA}$ with that of $V_{quant}$:
\beq
 \vert S_{quant}^2 \vert = {1 \over 2 \alpha} {\kappa^2 \over 16 \pi^2} M_p^2.
\label{sigmaf}
\eeq

The structure we have outlined above constitutes what is usually called {\it supersymmetric hybrid inflation}\cite{hybrid1,hybrid2,hybrid3,linderiotto,shafi1,shafi2,
shafi3,fischler}, but with the modification
$W_R$.   Again we see that with the assumption of small field excursions (compared
to $M_p$), and some modest assumptions about naturalness, hybrid inflation is {\it almost} inevitable.  After further studies of this structure, we will
subject the various assumptions to closer scrutiny, and ask whether some may be relaxed.

So far we have neglected $W_R$.  This term creates an additional potential on the pseudo moduli space (indeed it is not sensible to speak of such a moduli
space in anything but an approximate sense, even neglecting supersymmetry breaking).  The leading correction to the potential behaves as
\beq
\lambda \mu^2
S^N \over M_p^{N-2}.\eeq 
For sufficiently large field, this overwhelms both $V_{SUGRA}$ and $V_{quant}$.  The potential, in this regime, is not flat enough to inflate.
As we will see, this constrains $\mu$.

\section{Inflation in the Single Field model}
\label{kahlerconstraints}

In this section, we attempt to implement inflation in the single field model.  We will encounter difficulties, finding that the model  is not compatible
with facts of cosmology except for very large $N$.  But the model will be illustrative and is readily modified to accommodate astrophysical observations.

Let us first suppose that $W_R$ is sufficiently suppressed that it can be ignored during inflation.  We will
 quantify this in the next section.  Our interest is in inflation on the pseudo moduli space.
 The $S$
 potential arises from the Kahler potential.  
The slow roll conditions are:
\beq
\eta = {V^{\prime \prime} \over V}M_p^2 \ll 1;~~\epsilon = {1 \over 2} \left ({V^\prime \over V} \right )^2 M_p^2 \ll 1.
\eeq
If $V_{SUGRA}$ dominates,
both conditions are satisfied if $\alpha \ll 1$, and if $\vert S \vert \ll M_p$; these are minimal conditions for
successful inflation in any case.  

If there is a region where $V_{quant}$ dominates, i.e. if quantum corrections dominate over
the supergravity potential before reaching the ``waterfall regime" ($\vert \kappa t S \vert \gg 
vert \mu\vert)$, then
inflation may end before $S$ reaches the waterfall regime; $\eta \approx 1$ for
\cite{hybrid2}:
\beq
S_f^2 = {\kappa^2 \over 8 \pi^2} M_p^2.
\eeq
Alternatively, inflation may end when one enters the waterfall region,
\beq
S_{wf} = {1 \over \kappa} \mu 
\eeq
We will see now that the requirement that $\kappa$ be suitable to lead to sufficient inflation yields
\beq
S_{qu} \ll S_{wf}
\eeq
So inflation, if it occurs at all, takes place in the supergravity regime.  This is problematic, if nothing else, from the point of view
of the spectral index, $n_s$,
\beq
n_s = 1 + 2 \eta.
\eeq
In the supergravity regime, this is greater than one, which appears inconsistent with results from WMAP.\footnote{We should stress
that this is an issue in the particular class of hybrid inflation models considered here; "inflection point models" and
possibly other small field models can accommodate a ``blue" spectrum\cite{inflectionpoint}.  In the present context, we will shortly see that the spectrum
is blue as a result of quantum effects.}

Following \cite{linderiotto}, we treat $S$ as real (this implies no loss of generality provided
slow roll is valid and for small fields, i.e. when $W_R$ is negligible), and define
$\sigma = \sqrt{2} S$.
The number of $e$-foldings in the supergravity regime is:
\beq
{\cal N} = \int_{\sigma_{0}}^{\sigma_i}d\sigma {V \over V^\prime M_p^2}
\eeq
where $\sigma_0$ is the larger of $\sigma_{quant},\sigma_{wf}$.
This yields:
\beq
{\cal N} = { 1\over 2 \alpha} \log{\sigma_i \over \sigma_{quant}},
\eeq
The total number of $e$-foldings in the would-be quantum regime is
\beq
{\cal N}_{quant} = {1 \over 4 \alpha}.
\eeq
$\sigma_i$, the initial
value of the field, is not yet constrained by any
of our considerations.
For now we see that
significant inflation requires that $\alpha$ is small.  Generically, this is a tuning, of order $1 \over {\cal N}$ (possibly modulo a logarithm).  {\it This is irreducible}.

If there are $60$ or so $e$-foldings in the quantum regime, we can determine the values of the slow roll parameters purely in terms of known quantities.
$\delta \rho \over \rho$ is determined in terms of $\mu$ and $\kappa$:
\beq
{V^{3/2}/V^\prime} = 5.15 \times 10^{-4} M_p^3.
\eeq
This expression determines $\mu$ in terms of $\kappa$ (or vice versa):
\beq
\kappa = 0.17 \times \left ({\mu \over 10^{15} {\rm GeV}} \right )^2 = 7.1 \times 10^5  \times \left ({\mu \over M_p} \right )^2.
\label{kappaequation}
\eeq

Assuming $60$ e-foldings of inflation in the quantum regime, we have, then, for $S$ 60 e-foldings
before the end of inflation,
\beq
S_{60} = {\kappa \over 4 \pi} \sqrt{60} M_p.
\eeq
Substituting our result for $\kappa$, the condition that one not have already entered the waterfall regime is:
\beq
{S_{60} \over S_{wf}} = {\kappa^2 \sqrt{60} \over 4 \pi} \left ({M_p \over \mu} \right ) = 4 \times 10^{12} \left ({\mu \over M_p} \right )^3
\eeq
The requirement that this be much greater than one yields:
\beq
{\mu \over M_p} \gg 10^{-4}.
\label{mulimit}
\eeq
This is a rather high scale.  

We will see in section \ref{wrconstraints} that there is an {\it upper} bound on the scale of inflation, $\mu$ (depending on $N$).  
Only for very large $N$ or small $\lambda$ are scales as large as those of eqn. \ref{mulimit} achievable.

\section{Inflation in the Two Field Model}

The difficulty we have encountered in the single field model can be resolved by invoking the model of eqn. \ref{modifiedhybrid}, with two Hubble-mass fields.  In this model,
as we noted, the inflation is not the partner of the Goldstino.
The quantum potential is now:
\beq
\delta K_{quant}(S,S^\dagger, I, I^\dagger) = {\kappa^2 \over 16 \pi^2} S^\dagger S\log(I^\dagger I).
\eeq
The condition on $\kappa$ required to obtain a suitable fluctuation spectrum is essentially as before, with $S$ replaced
by $I$; similarly for the formula for the number of $e$-foldings.  But the condition that
$I_{qu} \gg I_{wf}$ is now much different, since
\beq
I_{wf} = {\kappa \mu^2 \over \lambda}
\eeq
and $\lambda$ can be of order one.  Indeed, in this model, inflation ends when $I$ is sufficiently small that $\eta    \approx 1$, which
occurs well before reaching the waterfall regime.


In the limit in which we study,
in which supergravity corrections are unimportant, $\eta$ and $n_s$  are universal, and $\epsilon$ is small.
\beq
\eta = -{1 \over {\cal N}}; ~n_s = {1 +2 \eta}.
\eeq
So one expects, quite generally, that if inflation occurs in the quantum regime, $n_s \approx 0.98$.  Again, we note that
if inflation occurs in the supergravity regime, one predicts $n_s >1$, which appears to be ruled out by current CMBR observations.

\section{Tuning and the Number of Light Fields}
\label{lightfields}

So far, we have taken the point of view that in a supersymmetric framework, fields light on the scale $H_0$ are natural.
However, tuning is required to have at least one field much lighter than $H_0$.  Here we note that the degree of tuning
grows with the number of light fields which can mix with the inflaton (with the same quantum numbers as the inflaton).

In the case of a single field, we have seen that successful inflation requires that the parameter $\alpha$ be less than
$1 \over{\cal N}$, the number of $e$-foldings.   In the context of a landscape (or just standard 't Hooft'ian notions of tuning)
this corresponds to a tuning of order $1 \over{\cal N}$.  More fields require more tuning.  In particular, suppose we have
a set of singlets, $S_i$, $i = 1,\dots N$, all with the same $R$ charge as the superpotential.  Then, by a field redefinition,
we can take, for the linear term in the superpotential,
\beq
W = \mu^2 S_1.
\eeq
We can take the Kahler potential to be:
\beq
K = S_i^\dagger S_i + {1 \over M_p^2} [\alpha (S_1^\dagger S_1)^2 +\left ( \alpha_i S_1^\dagger S_1 S_1^\dagger S_i + {\rm c.c.} \right )
+{1 \over M_p^2} \left ( \alpha_{ij} S_1^\dagger S_1 S_i^\dagger S_j  + {\rm c.c.} \right ).
\eeq

Our assumption in considering tuning is that all of the $\alpha$ parameters are naturally of order one.  In order that $S_1$, say, have mass
of order $1/{\cal N}$, it is necessary that $\alpha$ be of this order, while the real and imaginary parts of $\alpha_{i}$ be of order ${1 \over \sqrt{{\cal N}}}.$
This suggests a fine tuning of order $1 \over {\cal N}^N$.  Indeed, this indicates that fine tuning is minimized if the inflaton does indeed lie in a multiplet with
the gravitino.

In a somewhat different context (D-brane inflation), an analysis of the fine tuning required to obtain inflation with multiple
fields has appeared in \cite{mcallister}.  There, an effective action was written for a set of fields, and coefficients in the lagrangian
chosen at random.  The models there are different in a number of respects -- there is no low energy supersymmetry,
terms are included to higher orders in fields -- yet the scalings observed in a Monte Carlo analysis are similar.  We see that
in the framework of supersymmetric effective lagrangians, these estimates are very simple.

\section{Constraints from $W_R$}
\label{wrconstraints}

In the presence of $W_R$, the system has supersymmetric minima, satisfying
\beq
S^N = {2 \mu^2 M_p^{N-2} \over \lambda}~~~\phi=0.
\eeq
At large $S$, the potential includes terms
\beq
\delta V_R = 2 \lambda \mu^2 {S^N \over M_p^{N-2}}+ {\rm c.c.}
\eeq
If these terms dominate, the system will be driven towards the supersymmetric minimum.  So if we insist that the
 system is driven to the $R$ symmetric stationary point, we must require that these terms are small, and this in turn places limits on the scale
$\mu$ (or, through equation \ref{kappaequation}, the coupling $\kappa$), as well as $\sigma_i$.

The analysis of the previous section goes through provided that $\sigma_i > \sigma_{quant}$, and that $V^{''}_R(\sigma_i) \ll V^{''}_{quant}(\sigma_i)$.   The real constraint
on the underlying model comes from the requirement that $V^{''}_R(\sigma_{quant}) \ll V^{''}_{quant}(\sigma_{quant})$.  This translates into a restriction on $\mu$, or equivalently
$\kappa$:
\beq
\left({\mu \over 10^{15}GeV} \right)^{2N-6} \ll { 0.34(69)^{N-2} \alpha^{N/2} \over \lambda N (N-1)  } \times 10^{-6}
\label{mulimit}
\eeq
For $N=4-6$, this yields for the maximal scale:
\beq
N=4:~\mu \approx 1.2 \times 10^{11}( \alpha \times 100)~{\rm GeV} \\
N=5:~\mu \approx 1.6 \times 10^{13}( \alpha \times 100)^{5/8}~{\rm GeV} \\
N=6:~\mu \approx 8.0 \times 10^{13}( \alpha \times 100)^{1/2}~{\rm GeV}
\eeq
Even for larger $N$, the scale is not extremely large; e.g. for $N=12$, it is only of order $8 \times 10^{14}$.

For $N=3$, there is no choice of $\mu$ for which $W_R$ does not dominate.  One can try to resolve this by including higher order terms in the Kahler potential
and considering supergravity corrections to the {\it potential} of the form:
\beq
\delta V = \beta \mu^4 {\vert S \vert^4 \over M_p^2}.
\eeq
However, in this case, all of the activity occurs, for $\beta \sim 1$, for $S \sim \sqrt{\alpha} M_p$, which does not seem consistent with the idea of small field inflation,
unless $\alpha$ is tuned to be {\it extremely} small.  Alternatively, the coefficient of the operator appearing in $W_R$ might be very small.  Calling
\beq
W_R = {\lambda \over 2(N+1)} {S^{N+1} \over M_p^{N-1}},
\eeq
in the case
$N=3$, we require
\beq
\lambda \ll 10^{-5} \alpha^{3/2}.
\eeq
Indeed, we could consider, for any $N$, the possibility that $\mu$ is larger than implied by eqn. \ref{mulimit}, and $\lambda$ is small.  The general condition is:
\beq
\lambda \ll \left ({\mu \over M_p} \right )^{6-2N} \alpha^{N/2} {2(6.65 \times 10^{-5})^{N-2}\over N(N-1)}.
\eeq
Thus $\lambda$ has to be extremely small for reasonable values of $N$.  Small $\mu$ is arguably more plausible.

In the two-field model, one has similar constraints on the inflationary scale.  A coupling
like that of $W_R$, with $S^{N+1}$ replaced by $S I^N$, for example, has essentially identical effects.

Note that we now have now enumerated three types of constraints/tunings:
\begin{enumerate}
\item  $\alpha$ must be small, comparable, up to a logarithmic factor, to $1/{\cal N}$, one over the number of $e$-foldings.
\item  $\mu$ and $\kappa$ must be small, in order that the superpotential corrections not drive $S$ to a large field regime.
\item  The initial conditions for $S$ are constrained.  $S$ must lie in a range small enough that, at least for a time, the quantum
potential is dominant, and large enough that inflation can take place.
\end{enumerate}

\section{Reheating}
\label{reheating}

So far, we have not discussed the quantum numbers and couplings of $\phi$.  These are important since it is the lifetime of $\phi$
which determines the reheating temperature.  For the scales of interest here,
\beq
\phi = \kappa^{-1/2} \mu
\eeq
and this can readily be of order grand unified scales.  So many hybrid inflation models take $\phi$ to be, say, in the adjoint representation of some grand unified group.
If $S$ is a singlet, this is potentially problematic, since there may be additional light states (in the final vacuum); these can spoil unification
and lead to other difficulties.  A very simple possibility is to take $\phi$ to be
a gauge singlet, and couple $\phi$ to some charged fields; schematically
\beq
\delta W = \lambda \phi \bar 5 5.
\eeq
In the vacuum, the masses of $\bar 5$, $5$, are of order $m_Q = \lambda \phi$.  So the $\phi$ lifetime is of order
\beq
\Gamma = \left ({\alpha_s \over \pi} \right )^2 {\lambda^2 \over 4 \pi} {m_\phi^3 \over m_Q^2}.
\eeq
This can be rewritten, using $\phi = \kappa^{-1/2} \mu$, and the relation between $\kappa$ and $\mu$:
\beq
\Gamma = {1 \over 4 \pi} \left ( {\alpha_s \over \pi} \right )^2 \kappa^{5/2} \mu
\eeq
$$~~~~\approx 3 \times 10^{9} \left ( {\mu \over 10^{15} } \right )^6 ~{\rm GeV}.
$$
For $\mu = 10^{12}$ GeV, this corresponds to $\Gamma =3 \times 10^{-9}$ GeV, or a reheat temperature of order $10^{4.5}$ GeV.  Larger $\mu$ leads
to higher reheat temperatures.  Such temperatures are clearly interesting from the point of view of the gravitino problem and other cosmological issues.

\section{Non-Hybrid Scenarios}
\label{nonhybrid}

So far, we have insisted on an unbroken, discrete $R$ symmetry at the end of inflation.  But we might relax this.  For example, consider a model with superpotential
\beq
W = \mu^2 S - {\lambda \over 2(N+1)} {S^{N+1} \over M_p^{N-2}}.
\label{wnonhybrid}
\eeq
Here we might try to arrange that the field, during inflation, rolls towards the supersymmetric minimum at
\beq
S_0^N = {2 M_p^{N-2} \mu^2 \over \lambda}.
\eeq
As we will see, the conditions for slow roll inflation can be satisfied for a range of initial field values, and adequate e-foldings and fluctuation spectrum obtained.
Inflation ends as the field moves towards the minimum.  It is important, however, that the energy not be negative at the end of inflation, and this
requires a small constant in the superpotential.  This constant breaks the discrete R symmetry.  While small, this term is not a small perturbation.
In particular, it gives a large contribution to $V^{\prime \prime}$ at the point where $V^{\prime \prime} \sim H^2$.  This issue has been discussed
in \cite{mooij}, where possible additional tunings and/or additional degrees of freedom which might permit suitable inflation have been considered.

Even if one builds a successful model,
when inflation ends, $S$ oscillates about $S_0$.  It is important that there be a mechanism to dissipate the energy of oscillation.  This does require coupling
to additional fields.  For example, adding again a $5$ and $\bar 5$ of $SU(5)$:
\beq
\delta W = \kappa S \bar 5 5,
\eeq
$S$ can decay to gauge boson pairs.  Note, however, that for sufficiently small $S$, the quantum contributions to the potential will dominate
over those we have considered up to now.   For initial configurations in this regime, the field will flow towards the origin.  This places a {\it lower} limit
on $\sigma_i$, which depends on $\kappa$.
Overall, then, the inflationary scenario we have outlined in the earlier sections seems less tuned and simpler than the non-hybrid scenarios.

\section{Supersymmetry Breaking}
\label{susybreaking}

The basic structure of the hybrid inflation superpotential is reminiscent of O'Raifeartaigh models.  It is interesting to include additional degrees of freedom so that $S$ is part
of a sector responsible for (observed) supersymmetry breaking.  For small $S$, in addition to the field $\phi$, we can include another field, $X$, with $R$ charge two and superpotential:
\beq
W = S (\kappa \phi^2 - \mu^2) +m X \phi.
\label{oraifeartaigh}
\eeq
Assuming that the scale of inflation is large compared to the scale of supersymmetry
breaking leads to consideration of the limit $\vert m \vert^2 \ll \vert \kappa  \mu^2 \vert$.
In this limit, the vacuum state has $\phi \approx \kappa^{-1/2} \mu$, $X = S = 0$, and
\beq
F_X \approx \kappa^{-1/2} m \mu.
\eeq
By choosing $m$, we can arrange $F_X$ as we please.  We might worry that $X$ is light, and relatively long-lived.  Classically, the $X$ mass vanishes.
Quantum mechanically, it is of order
\beq
m_X^2 = {\kappa^4 \over 16 \pi^2}{ m^4 \over m_S^6} F_X^\dagger F_X.
\eeq
Calling $m^2 = \epsilon \kappa \mu^2$ (we require $\epsilon < 1$ in order that $\phi$ have a vev), and noting $m_s^2 = \kappa \mu^2, ~F_X = \epsilon^{1/2} \mu^2$, we have
\beq
m_X^2 = {\epsilon^3 \over 16 \pi^2} \mu^2.
\eeq
So $\epsilon$ can be quite small, with $X$ still in the TeV range.

In the model as it stands, the $R$ symmetry is unbroken by loop effects.  This can be avoided through additional, ``retrofitted" couplings\cite{dinekehayias}, or through models like that of
\cite{shih}.
\section{Hybrid Inflation and Dynamical Supersymmetry Breaking}
\label{retrofitting}

It would be appealing if the scales appearing in our models of inflation (and supersymmetry breaking) could be understood dynamically.  The simplest implementation
of (metastable) dynamical supersymmetry breaking is through ``retrofitting"\cite{dfs,dinekehayias,kehayias}.  In particular, we want to generate the scales
$\mu^2$ and $m$, of eqn. \ref{oraifeartaigh} dynamically.  In these references, the scale $\mu^2$ was of order $\Lambda^3 \over M_p$, where $\Lambda$ is the scale
of some underlying, supersymmetry preserving but $R$ symmetry breaking dynamics.  In the present case, however, this would lead to too large a value of
$W_0$, the value of the superpotential at the end of inflation.  Instead, we follow \cite{dinekehayias,kehayias}, and consider theories with order parameters of dimension
one as well as dimension three.  These are singlet fields, which we will denote by $\Phi,~\langle \Phi\rangle \sim \Lambda$.  The $\Phi$ fields have mass of order $\Lambda$.  We replace
the superpotential of eqn. \ref{oraifeartaigh} by
\beq
W = S(\kappa \phi^2 - \lambda \Phi^2) + {\lambda^\prime \Phi^2 \over M_p} \phi Y.
\eeq
$\lambda$ need not be particularly small in order not to appreciably perturb the $\Phi$ dynamics, i.e. to satisfy the requirement
that $\Phi$ is massive compared to $S$, $\Phi$.  Then at scales well below $\Lambda$, the theory is that of eqn. \ref{oraifeartaigh}.  $<W> \sim \Lambda^3 \sim \mu^3$ at the
minimum of the potential, so the supergravity corrections to the potential are negligible at low energies.  In fact, we have
\beq
F_Y \approx m \phi = \Lambda^3 \kappa^{-1/2}/M_p
\eeq

\section{Conclusions}

So we have seen that small field inflation is likely to require supersymmetry, and that conventional
notions of naturalness also lead to the inevitable requirement of an $R$ symmetry.  This leaves two classes of models:  hybrid and
RBI.  In the former, we have seen that the requirement that the $R$ symmetry be discrete places an {\it upper} bound on the scale
of inflation, which makes observations of tensor modes in the CMB extremely unlikely.  Also inevitably, $n_s <1$, typically about $0.98$.
In the RBI case, there are also constraints on scales, and one requires some sort of soft breaking of the $R$ symmetry,
describable, perhaps, by spurion fields.

We have seen that there is a simple effective field theory description of these types of inflation, and that one can use the language of global
supersymmetry, perturbed slightly (but in critically important ways) by coupling to supergravity.  In this framework, the simplest models
have an inflaton which lies in a supermultiplet with the gravitino, but this is not necessary.  Indeed, we have seen that
once one considers higher dimension operators, there is an upper bound on the energy scale of inflation, and that
models with at least one additional field more readily lead to successful inflation.  One could go further than we have here in applying
the language of \cite{komargodskiseiberg} to this problem.

We have made much of the naturalness of supersymmetry for the problem of small field inflation, and one might wonder whether, in, say,
an anthropic landscape framework, these considerations would be relevant.  Needless to say, it is hard to make a definitive statement;
it could be that supersymmetric states are very rare in the landscape, and that this overwhelms any tuning considerations (just as for the
Higgs\cite{dinethomas}).  But we can ask the question a different way.  Given the assumption of a supersymmetric landscape, could it be that the requirement of inflation
accounts for a little (or not so little) hierarchy?  E.g. small $\kappa$ and correspondingly small $\mu$ might be disfavored by landscape distributions;
depending on correlations between $m$ (in eqn. \ref{oraifeartaigh}) and $\mu$, one might be driven to larger scales of supersymmetry than might be
expected from naive naturalness considerations.  This issue will be discussed elsewhere\cite{dinepack}.

We have not discussed the problem of initial conditions at any length.  Certainly there are constraints on the initial values of the fields,
their velocities, and the degree of homogeneity required.  These issues look different in different contexts, and we will leave them for further
study.  One striking feature of this framework is the nature of the cosmological moduli problem.  In the models discussed here, the minimum of the potential
is a point with an approximate $R$ symmetry.  Moduli which are charged under the symmetry naturally sit near the origin, and are not particularly light.
Neutral fields (such as the field $X$ in the retrofitted models) naturally sit near the origin as a result of the {\it accidental, continuous} $R$ symmetry.
There seems to be no moduli problem in this context.  This issues will be explored further elsewhere.

Finally, as we noted at the beginning, almost by definition it is hard to make general statements about large field inflation.  One could attempt
this as a limit of the analysis we described above, but then the effective lagrangian has many more ``relevant" parameters, and one can
probably, at best, simply state constraints consistent with observations on combinations of these.  It is hard to see how predictions can emerge without a detailed
microscopic understanding of the underlying gravity (supergravity) theory.

\noindent
{\bf  Acknowledgements:}  Conversations with Willy Fischler, Dan Green, and Liam McAllister helped clarify our thinking about many issues here.
Correspondence with David Lyth provided valuable perspective.
M.~Dine thanks Stanford University and the Stanford Institute for Theoretical
Physics for hospitality during the course of this project.  He thanks the Theory Group at the University
of California, Berkeley and the Theory Group at Lawrence Berkeley National Laboratory for their hospitality as well.  We thank the authors
of ref. \cite{mooij}, and especially S. Mooij and M. Postma, for conversations and pointing out an error in our original discussion of
the non-hybrid models of section \ref{nonhybrid}.
This work was supported in part by the U.S.~Department of Energy.

\bibliography{hybridrefs}{}
\bibliographystyle{unsrt}
\end{document}